\documentclass[12pt,preprint]{aastex}

\usepackage{graphicx}

\shorttitle{Zero-frequency magnetic fluctuations in homogeneous cosmic plasma revisited}
\shortauthors{F. Caruso \& V. Oguri}

\begin{document}

\title{Zero-frequency magnetic fluctuations \\
in homogeneous cosmic plasma revisited}

\author{F. Caruso}
\affil{Centro Brasileiro de Pesquisas F\'{\i}sicas \\
Rua Dr. Xavier Sigaud 150, 22290-180, Rio de Janeiro, RJ, Brazil}
\email{francisco.caruso@gmail.com}

\author{and V. Oguri} 
\affil{ Universidade do Estado do
Rio de Janeiro  \\
Rua S\~ao Francisco Xavier 524, 20550-013, Rio de Janeiro, RJ, Brazil}
\email{oguri@uerj.br}




\begin{abstract}
Magnetic fluctuations in a non-magnetized gaseous plasma is revisited and calculated without approximations, based on the fluctuation-dissipation theorem. It is argued that the present results are qualitative and quantitative different form previous one based on the same theorem. In particular, it is shown that it is not correct that the spectral intensity does not vary sensitively with $k_{cut}$. Also the simultaneous dependence of this intensity on the plasma and on the collisional frequencies are discussed.

\end{abstract}

\section{Introduction}

Fluctuations of physical quantities near zero frequency have been investigated by several authors since the papers of \citet{Johnson} and \cite{Nyquist}. A general theory on the fluctuation-dissipation theorem, which will be the starting point of this paper, was developed by \citet{Kubo}. To the best of our knowledge, a concrete expression for the low-frequency spectrum of fluctuations of magnetic fields in a thermal plasma was obtained for the first time by \citet{Tajima}. They found a peak around $\omega =0$ magnetic fluctuation which was interpreted as the evanescent energy component of electromagnetic fluctuations ``screened'' in a plasma below the plasma frequency. The impact of such result into the cosmic microwave background was then investigated by \citet{Tajima-Shibata}. Although in these two references the authors claim that the fluctuations were rigorously computed, several approximations were indeed made and they were not able to get an unique formula covering the low- and high-frequency spectrum. The aim of this paper is to reevaluate the derivation of the spectrum of magnetic fluctuations avoiding approximations in the low-frequency region and also in the transition between the low- and the high-frequency spectrum. Several different behaviors between our results and the previous one, mainly in the low-frequency part of the spectrum, are found and discussed.

\section{The first predictions}\label{first_prediction}

The fluctuation-dissipation theorem developed by \citet{Kubo} is able to deal with the thermal fluctuations inside a plasma in or near thermal equilibrium. The expression for the magnetic field fluctuation  in an homogeneous isotropic non-magnetized equilibrium plasma was obtained by \citet{Tajima} looking at waves in such a plasma. In an electron-positron plasma, for example, the magnetic fluctuations in wave number and frequency space is given by

\begin{eqnarray}\label{original-formula}
\nonumber\frac{\langle B^2\rangle_{\vec k, \omega}}{8\pi} &=& \frac{2\hbar \omega}{e^{\hbar\omega/k_{_B}T} - 1}\, \eta \omega_p^2\, \times\\
\ \\
\nonumber &\times& \frac{k^2c^2}{(\omega^2 + \eta^2)k^4c^4 + 2\omega^2 (\omega_p^2 - \omega^2 - \eta^2)k^2c^2 + [(\omega^2 - \omega_p^2)^2  + \eta^2 \omega^2] \omega^2}
\end{eqnarray}
where $k_{_B}$ is the Boltzmann constant and $\omega_p$ and $\eta$ are, respectively, the plasma and the collisional frequencies. Here $\omega_p^2 = \omega_{p\, e^-}^2 + \omega_{p\, e^+}^2$, and
$$\omega_p = \sqrt{\frac{n 4 \pi Z e^2}{\gamma m}}$$
being $n$ the particle density inside the plasma, $Z$ is the atomic number of these constituents, $e$ and $m$ are respectively the charge and the mass of the electron, and $\gamma$ is the Lorentz factor given by $\gamma = (1 + kT/mc^2)$.
This result can be integrated in $\mbox{d}^3\vec k = 4\pi k^2\mbox{d}k$ to get (the Fourier transform)
$$\frac{\langle B^2\rangle_{\omega}}{8\pi} = \int \frac{\mbox{d}\vec k}{(2\pi)^3}\, \frac{\langle B^2\rangle_{\vec k, \omega}}{8\pi}$$
and the magnetic energy density is $\displaystyle \int_{-\infty}^{\infty} (\mbox{d}\omega/2\pi)(\langle B^2\rangle_{\omega}/8\pi)$.

The integral of eq.~(\ref{original-formula}) over wave numbers shows a high wave number divergence. According to \citet{Tajima}, this is expected since the derivation is based on classical fluid equations of motion and the constant collision frequency $\eta$ is independent of $k$. However, they prefer to carry on their analyzes in the simpler phenomenological approach. In order to overcome the large $k$ dependence, they first take the limit $\eta \rightarrow 0$ and then they integrate over $k$ to infinity, which corresponds to the vanishing cross section of collisions as $k \rightarrow \infty$. This is a very delicate point and we will turn back to this point in Section~\ref{general}. For both the high frequency and high wave number limits the authors emphasized that the expression of eq.~(\ref{original-formula}) has a substantial value only where $\omega^2 -c^2k^2 - \omega_p^2 \simeq 0$. The combined high-frequency and high wave-number limits were get by letting $\eta \rightarrow 0$. The expression for the low-frequency spectrum was obtained by breaking up the $k$ integral into two intervals, by introducing a cutoff value $k_{\mbox{\scriptsize cut}}$, with $x_{\mbox{\scriptsize cut}} \equiv k_{\mbox{\scriptsize cut}}c/\omega_{p e}$. In the integration from 0 to $k_{\mbox{\scriptsize cut}}$, $\eta$ was kept finite while in the integral from $k_{\mbox{\scriptsize cut}}$ it was used the approximation $\eta \rightarrow 0$. The expressions obtained for the high and low parts of the spectrum was, respectively:

\begin{equation}\label{tajima_high}
\frac{\langle B^2\rangle_{\omega}}{8\pi} = \frac{T}{2\pi} \delta(\omega) \int \frac{\omega_p^2}{\omega_p^2 + c^2k^2}\, k^2 \mbox{d}k
+ \frac{1}{2\pi c^3}\, \frac{\hbar}{e^{\hbar\omega/k_{_B}T} - 1}\, (\omega^2 - \omega_p^2)^{3/2}
\end{equation}
and
\begin{eqnarray}\label{tajima_low}
\nonumber\frac{\langle B^2\rangle_{\omega}}{8\pi} &=& \frac{1}{\pi^2}\, \frac{\hbar \omega^\prime}{e^{(\hbar\omega^\prime_{pe}/k_{_B}T)\omega^\prime} - 1}\,  2\eta^\prime \left(\frac{\omega_{pe}}{c}\right)^3 \times \int \frac{x^4}{(\omega^{\prime 2} + \eta^{\prime 2}) x^4+ \cdots}\,\mbox{d}x +  \\
\ \\
\nonumber &+& \frac{\hbar (\omega^{\prime 2} - \omega_p^{\prime 2})^{3/2}}{2\pi e^{(\hbar\omega_{pe}/k_{_B}T)\omega^\prime} - 1}\, \left(\frac{\omega_{pe}}{c}\right)^3 \times \Theta (\omega - \sqrt{c^2k_{\mbox{\scriptsize cut}}^2 + \omega_p^2})
\end{eqnarray}
where $\Theta$ is the Heaviside step function, $\eta^\prime \equiv \eta/\omega_{pe}$, $\omega^\prime \equiv \omega/\omega_{pe}$, and $\omega_p^\prime \equiv \omega_p/\omega_{pe}$.

Finally the zero frequency limit of the magnetic fluctuations is give by
\begin{equation}\label{zero_lim_tajima}
\lim_{\omega\rightarrow 0} \frac{\langle B^2\rangle_{\omega}}{8\pi} = \frac{\hbar \omega^\prime}{\pi^2(e^{\hbar\omega_{pe}\omega^\prime/k_{_B}T} - 1)}\, 2 \left(\frac{\omega_{pe}}{c}\right)^3\, \frac{1}{\eta^\prime} \int_0^{x_{\mbox{\scriptsize cut}}}\, \mbox{d}x
\end{equation}

At this point the frequency spectral intensity was plotted for a temperature $T=~10^{10}$~K by requiring that the value of $k_{\mbox{\scriptsize cut}}$ (or $x_{\mbox{\scriptsize cut}}$) provide a smooth behavior at the joint between the low-frequency spectrum and the black-body spectrum. The choice was  $k_{\mbox{\scriptsize cut}} \sim \omega_{pe}/c$ or ($x_{\mbox{\scriptsize cut}} \sim 1)$. The result for other values of the temperature were presented in another paper by \citet{Tajima-Shibata}. The main claims by these authors was that the intensity of the spectrum does not vary sensitively with $k_{\mbox{\scriptsize cut}}$ and that near $\omega = 0$ the spectrum goes like $\omega^{-2}$. Let us now show our results.

\section{General Result}\label{general}

We have integrated eq.~(\ref{original-formula}) over $k$ analytically, without any approximation, by the partial fractions technique. The exact result for the indefinite integral over the wave number is:

\begin{eqnarray}\label{indef_integ}
\nonumber S(\omega) \equiv \frac{\langle B^2\rangle_{\omega}}{8\pi} &=& u_\omega b y + \frac{u_\omega}{2} r^{3/2} \left\{ \frac{1}{2} \sin \frac{3\theta}{2} \ln \left| \frac{R_{-}(y)}{R_{+}(y)}\right| \right.+ \\
\ \\
\nonumber&\,& \qquad + \left. \cos\frac{3\theta}{2} \left[\arctan\left(\frac{y-a^\prime}{b^\prime} \right) + \arctan\left(\frac{y+a^\prime}{b^\prime} \right) \right]\right\}
\end{eqnarray}

\noindent with the following definitions: $\ell_{_D} \equiv c/\omega_p$ is the Debye length, $x \equiv \omega/\omega_p$, $y \equiv k/k_0 = kc/\omega$, $\eta^\prime = \eta T/\omega_p$,
$a \equiv 1 - (x^2+ \eta^\prime)^{-1}$, $b \equiv (\eta^\prime/x)(x^2+ \eta^\prime)^{-1} $, $R_{\pm} (y) = y^2 \pm 2 a^\prime y + r^\prime$, $a^\prime = \sqrt{(r+a)/2}$, $b^\prime = \sqrt{(r-a)/2}$, $r = (a^2+b^2)^{1/2}$, $\sin (3\theta/2) = (ba^\prime + a b^\prime) r^{-3/2}$, $\cos (3\theta/2) = (aa^\prime - b b^\prime) r^{-3/2}$, and

\begin{equation}
u_\omega \equiv \frac{\hbar/\pi^2}{e^{\hbar \omega/k_{_B}T}-1}\, \frac{\omega^3}{c^3} = \frac{\hbar/\pi^2}{e^{x/x_p}-1}\, \frac{x^3}{\ell_{_D}^3}
\end{equation}
where $x_p \equiv k_{_B}T/\hbar\omega_p$.

\begin{figure}[!htbp]
{\includegraphics[height=6.cm]{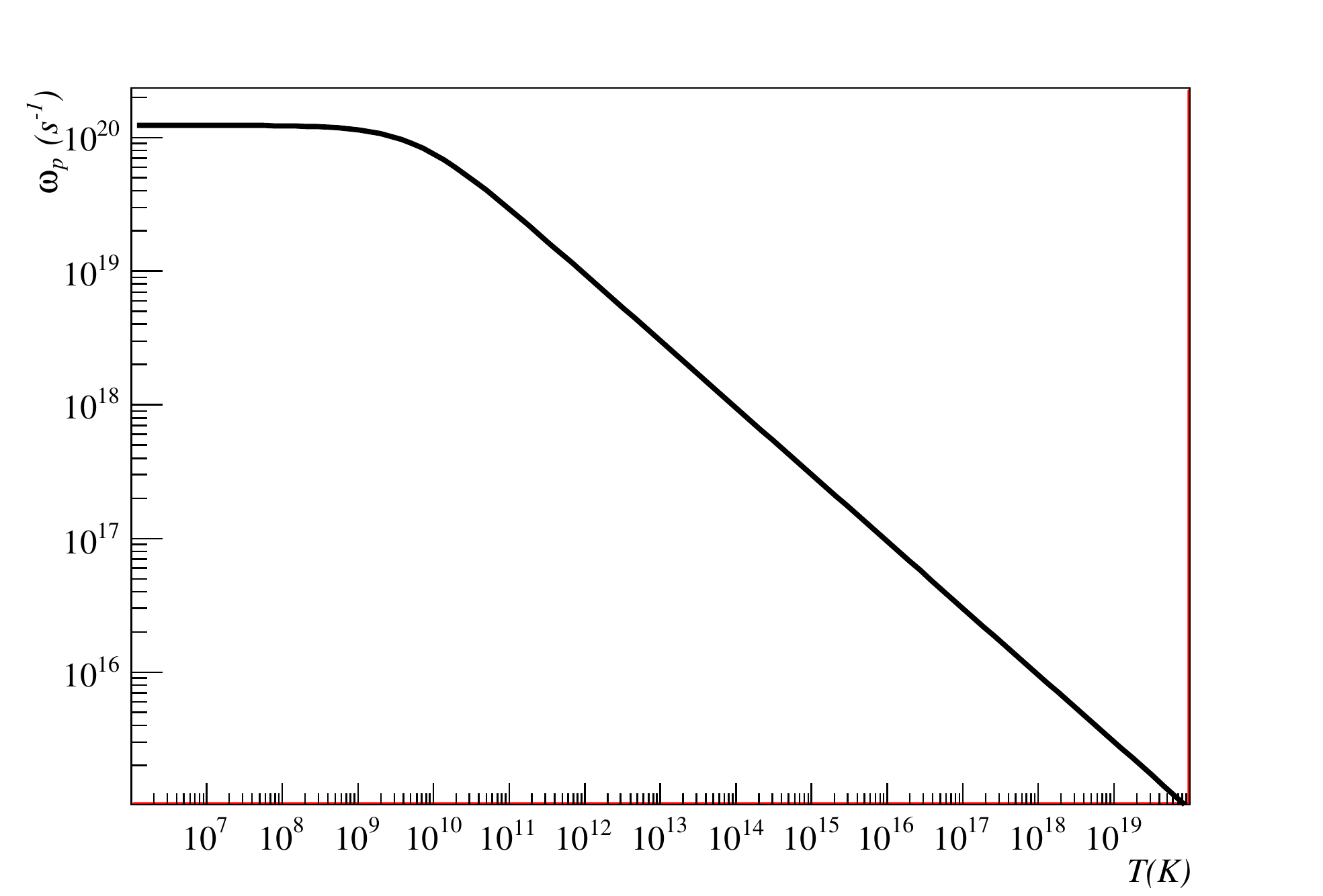}}%

\vspace*{-6.cm}
\hspace*{8cm}
{\includegraphics[height=6.cm]{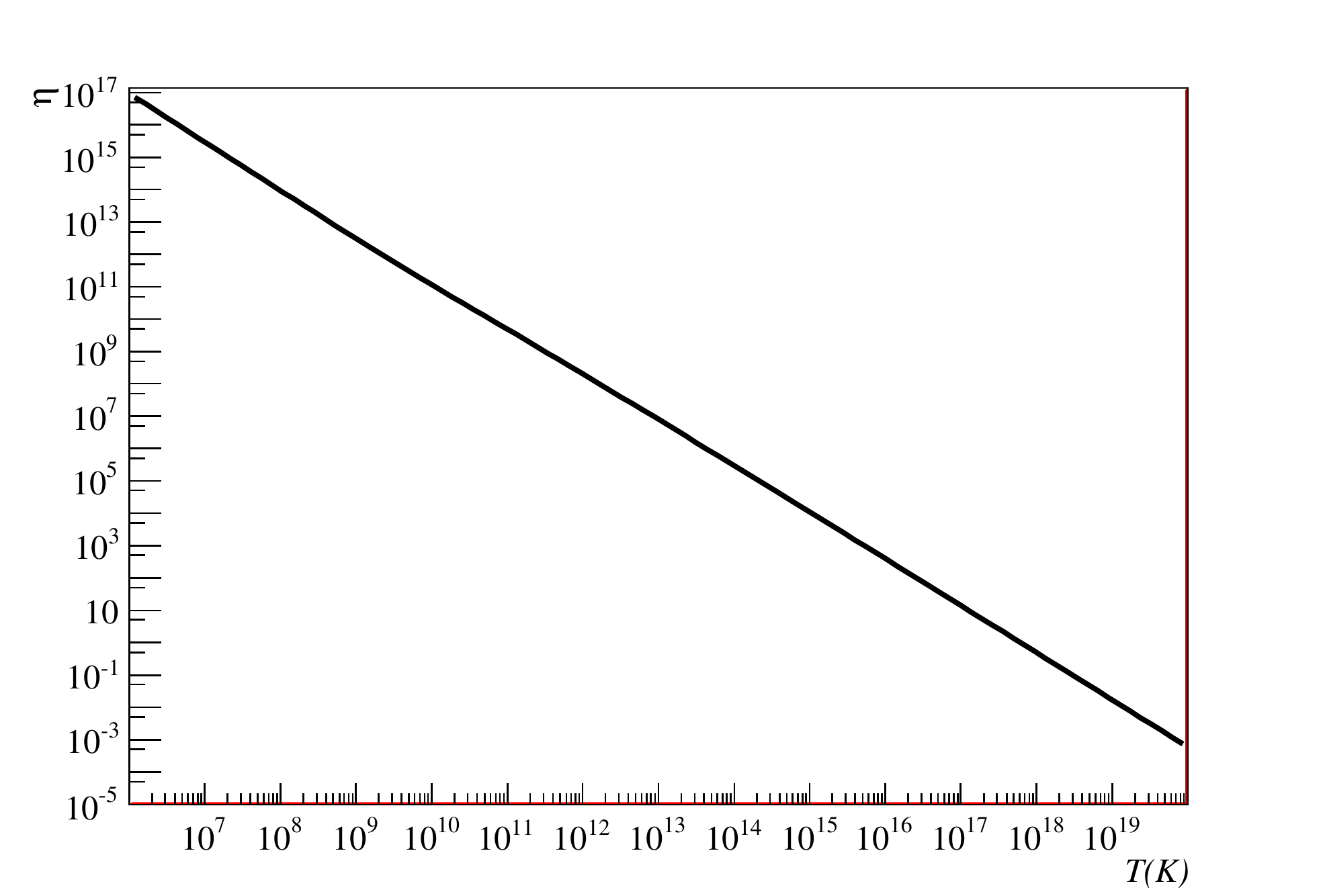}}%
\caption{(a) Plasma frequency and (b) plasma collison frequency, both as a function of plasma temperature.}
\label{fig-omega}
\end{figure}

Note that the general result shows only a linear divergence in $k$ restricted to the first term of eq.~(\ref{indef_integ}). This term, however, cannot be simply discarded by doing the limit $\eta\rightarrow 0$ before we integrate over $k$, as did by \citet{Tajima}, since it plays a very important role when the strict limit $\omega/\omega_p \rightarrow 0$ (which we prefer to indicate from now on by $\omega \ll \omega_p$) is to be considered, even when $k$ is large. Indeed, if we discard it for all large values of $k$ it can be shown that the limit $x\ll 1$ of $S(\omega)$ will be negative. Therefore, our result for the definite integral can be put in the form
\begin{equation}\label{our-result}
S(\omega) = S_1 + S_2 + S_3
\end{equation}
where
\begin{equation}\label{our-limit}
S_1 = u_\omega b \int_0^\infty \mbox{d} y; \qquad S_2 = \frac{\pi u_\omega}{2}\, (a^\prime b + a b^\prime); \qquad  S_3 = \frac{\pi u_\omega}{2}\, (a a^\prime - bb^\prime)
\end{equation}
The term $S_1$ will be taken as $S_1 = u_\omega b\, y_{\mbox{\scriptsize cut}}$ with $y_{\mbox{\scriptsize cut}}$ as large as we want. This will render the confrontation with the previous result easier. We note that for $k > 10^{15}$~cm$^{-1}$, $S_2 \rightarrow 0$. These three terms can be written as functions of the original parameters, considering that

$$ a a^\prime = \frac{1}{\sqrt{2}} \frac{\omega^2 +\eta^2 - \omega_p^2}{\omega^{1/2}(\omega^2 +\eta^2)^{3/2}} \left[\sqrt{(\omega^2 +\eta^2) [(\omega^2-\omega_p^2)^2 + \eta^2 \omega^2]} + \omega(\omega^2 + \eta^2 -\omega_p^2)\right]^{1/2}$$
and
$$ b b^\prime = \frac{1}{\sqrt{2}}  \frac{\eta\omega_p^2}{\omega^{3/2}(\omega^2 +\eta^2)^{3/2}} \left[\sqrt{(\omega^2 +\eta^2) [(\omega^2-\omega_p^2)^2 + \eta^2 \omega^2]} - \omega (\omega^2 + \eta^2 -\omega_p^2)\right]^{1/2}$$
Compared to eqs.~(\ref{tajima_high})-(\ref{zero_lim_tajima}) it is immediately evident how our result is different from those of equations,
showing a much more complicated dependence of the frequency spectrum on the variable $\omega$, and on the parameters $\omega_p$ and $\eta$,
which dependencies on plasma temperature are shown, respectively, in Figures~\ref{fig-omega}.

\begin{figure}[hbtp]
\centerline{\includegraphics[height=9.cm]{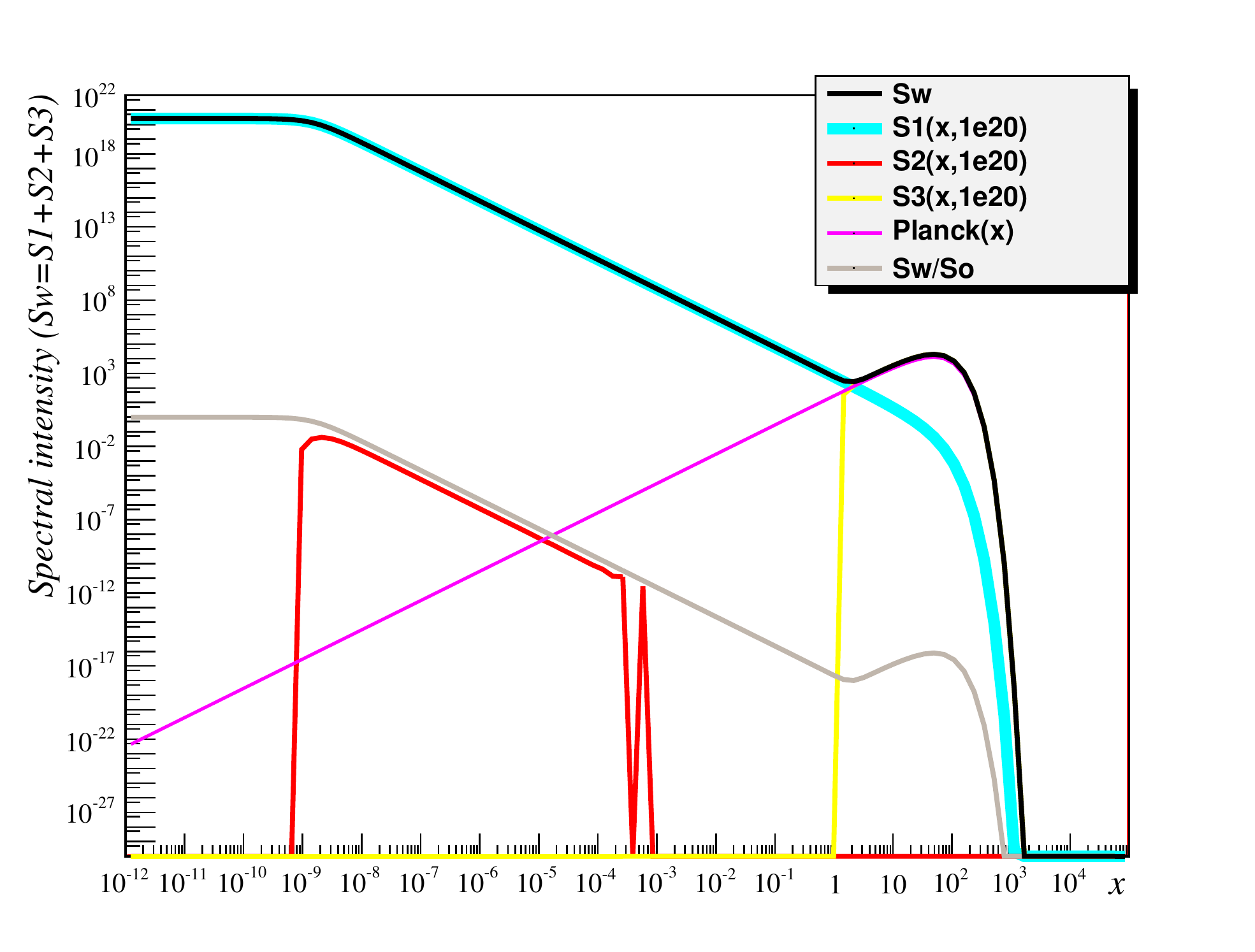}}%
\caption{Non normalized $S(\omega)$ and normalized $(S(\omega)/S_\circ)$ spectral intensities showing the different components of the spectrum, $S_1$, $S_2$ and $S_3$, as given by eq.~(\ref{our-limit}).}
\label{fig-full}
\end{figure}

Our full result $S(\omega)$ is shown in Fig.~\ref{fig-full}, where both the normalized (gray curve) and non normalized (green curve) spectral intensities are
given, for $k_{cut} = 10^{20}$~cm$^{-1}$ e $T = 10^{10}$~K.

In this figure we have also plotted each one of the terms that
contribute to $S(\omega)$ are shown in different colors. The deep we see in this figure near $x=1$ tends to disappear as $k_{cut}$ goes to high values; as $k_{cut}$ decreases up to $\simeq 10^{11}$~cm$^{-1}$, the ordinate of the deep tends to zero.

\begin{figure}[!htbp]
\centerline{\includegraphics[height=8.0cm]{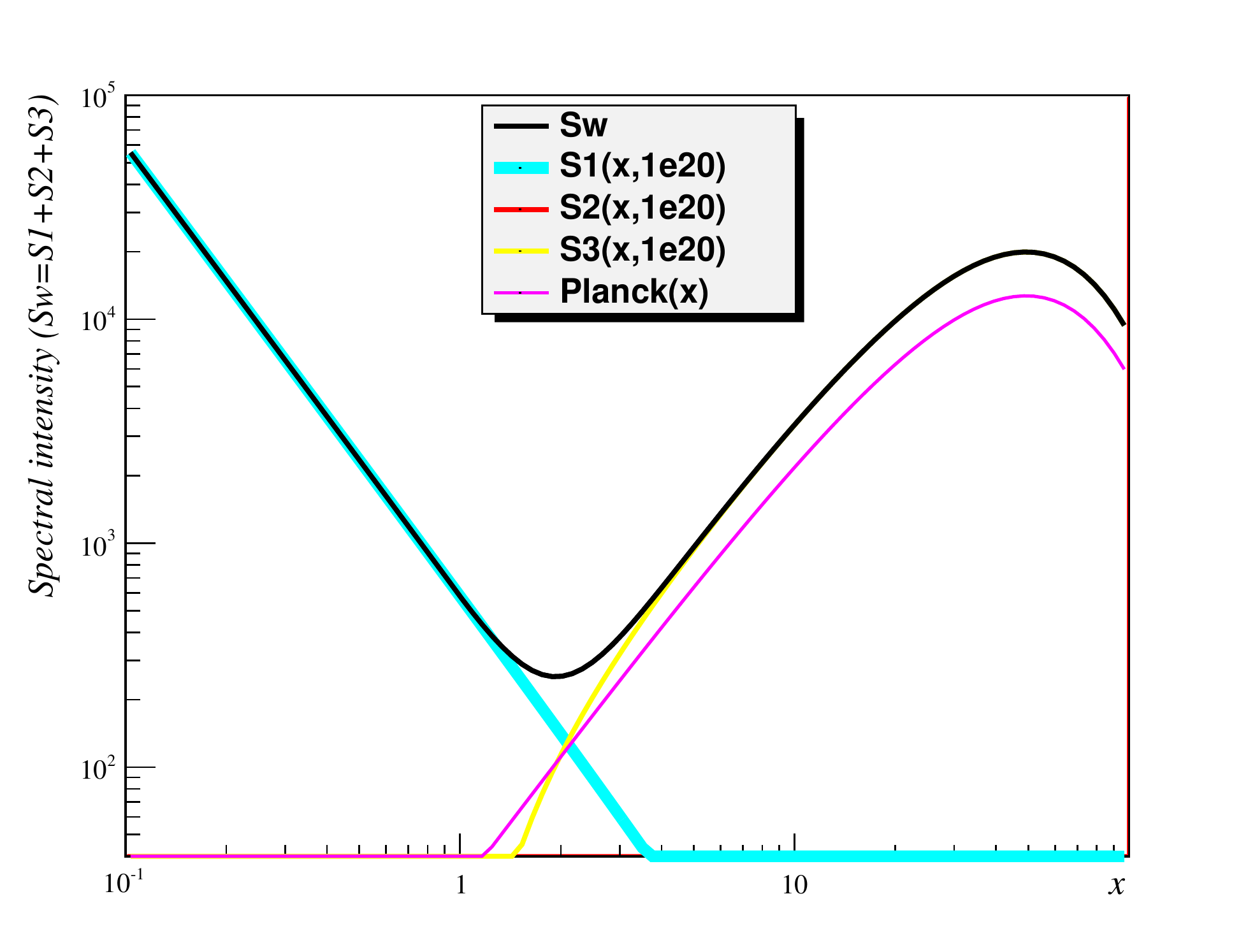}}%
\caption{Detail of the spectral intensity near $\omega \simeq \omega_p$.}
\label{fig-detail}
\end{figure}

The detail of the non normalized spectral intensity near $\omega \simeq \omega_p$ is shown in Fig.~\ref{fig-detail}. It shows naturally a smooth behavior between the low-frequency spectrum and the blackbody spectrum, which is constructed by hand in \cite{Tajima-Shibata}.

 If we compare this graph to the correspondent one shown in Fig.~1(a) of the paper of \cite{Tajima-Shibata}, we see that our result, for $x> 1$, is a classical blackbody radiation spectrum which goes to zero at $x \simeq 10^3$ while their blackbody spectrum has a much greater width (more than two orders of magnitude).

\begin{figure}[!htbp]
\centerline{\includegraphics[height=7.cm]{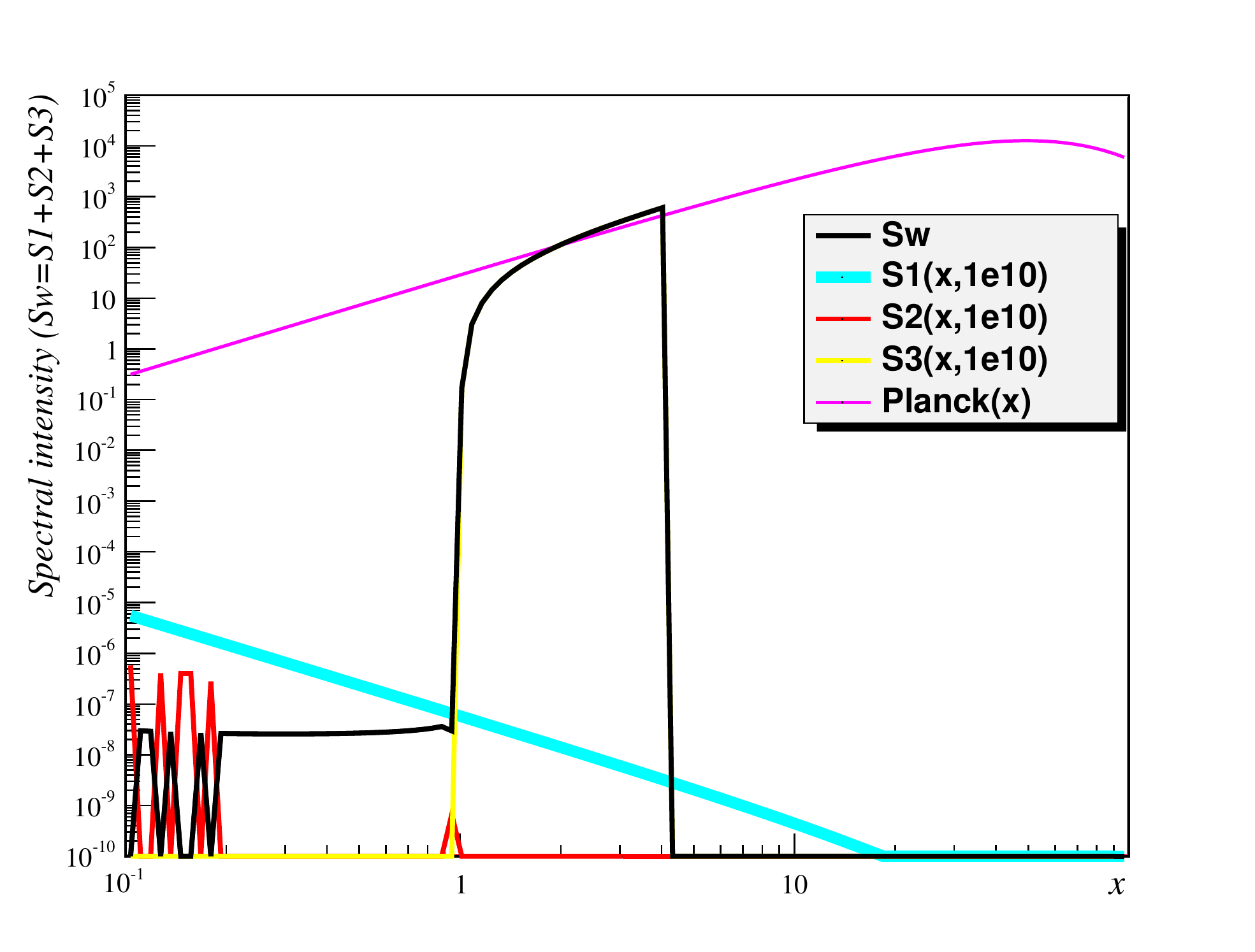}}%
\caption{Detail of the spectral intensity fluctuations for $k_{cut} = 10^{10}$~cm$^{-1}$.}
\label{fig-detail-k10}
\end{figure}

Another important qualitative difference between the non approximate and the approximate results is that we found a very peculiar oscillations in $S(\omega)$ for the $x\simeq 10^{-1}$ region of the spectrum, as can be seen from Figure~\ref{fig-detail-k10}. These oscillations occur in an $x$ region where the classical blackbody spectrum still have a significant value; there is however, in this case, a strong interference in the total spectrum $S(\omega)$, eq.~(\ref{our-result}), due to a change of sign of the function $S_3(\omega)$. Such a kind of behavior was found just for cut-off values of the order of  $k_{cut} = 10^{10}$~cm$^{-1}$. For values of $k_{cut}$ greater than this one such fluctuations disappear. In any case, this feature confirm our statement that the result can vary sensitively with $k_{cut}$, contrary to what was sustain by \cite{Tajima-Shibata}.

\begin{figure}[!htbp]
\centerline{\includegraphics[height=7.5cm]{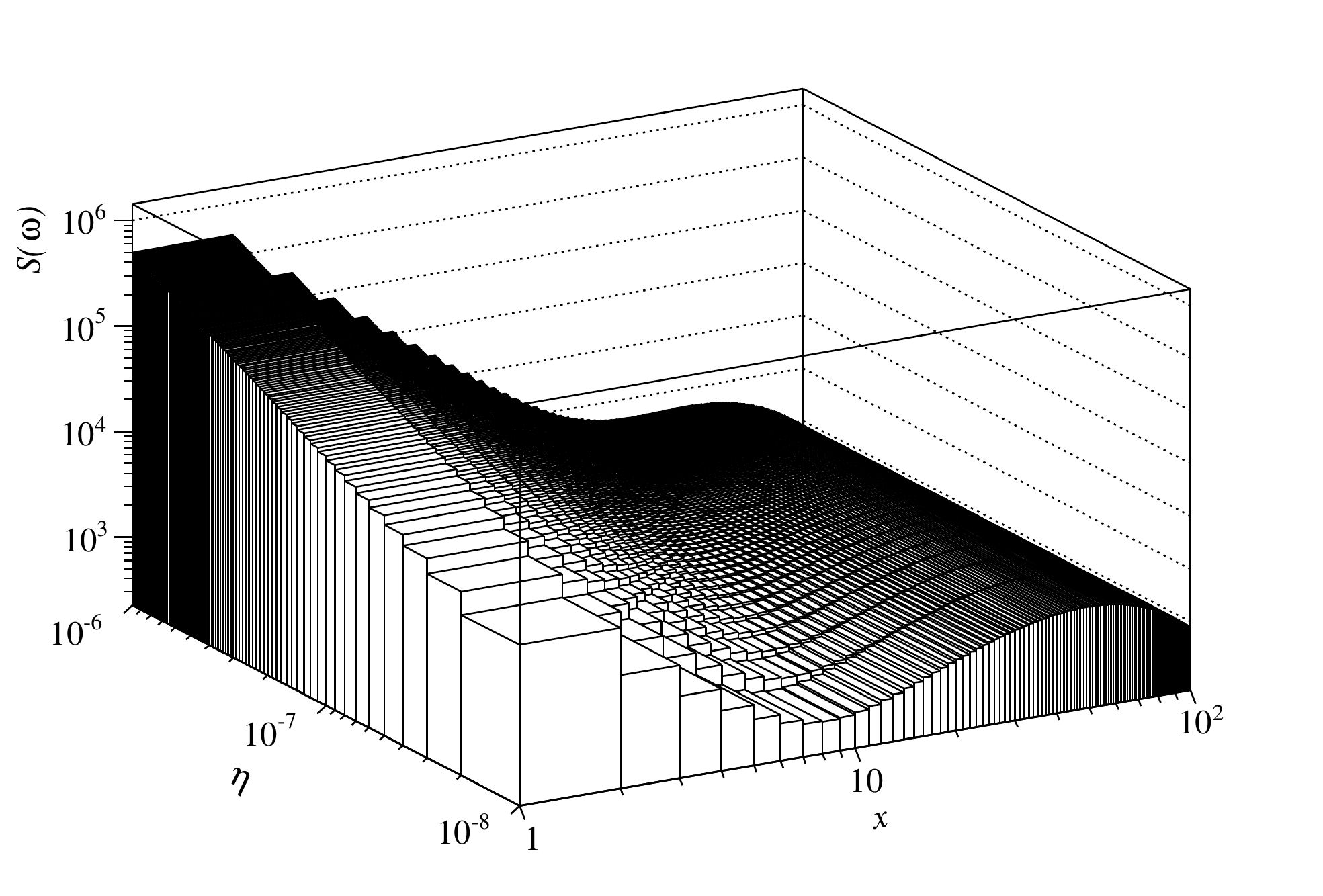}}%
\caption{Dependence of the spectral intensity on $\eta$ and $\omega$.}
\label{fig-surf6}
\end{figure}

Finally, we have studied the behavior of $S(\omega, \eta)$ by varying $\omega$ e $\eta$. The result is shown in Figure~\ref{fig-surf6}. Notice that just the peak of
the zero-frequency plasma spectrum depends on $\eta$ (the blackbody part remains unchanged). Indeed, the spectral intensity varies two orders of magnitude by
varying $\eta$ by two orders too, namely, it goes from $S(\omega, \eta) \simeq 10^6$, for $\eta \simeq 10^{-6}$, to $\simeq 10^4$, for $\eta \simeq 10^{-8}$. Thus, our result indicates that, when one goes backwards in time, temperature grows, dynamo action is enhanced (since $\eta$ goes down), and the resonance peak of the zero-frequency plasma peak goes down.

\section{Discussions}\label{disc}

In this paper we have computed the spectrum of magnetic fluctuations of an homogeneous cosmic plasma avoiding any approximations. Several different behaviors between our results and the previous one obtained by \cite{Tajima-Shibata}, mainly in the low-frequency part of the spectrum, are found and discussed. It is important to stress that the exact result indicates that the peak of the zero-frequency spectrum can indeed vary sensitively with the cut-off value $(k_{cut})$.

In the light of this new result, and following the papers of \cite{Tajima-Shibata} and \cite{Caruso}, the problem of establishing an upper limit for fractal space dimensionality from COBE data can be revisited.

Our results can still be improved towards cosmological applications by computing the Fourier transformed volume element $\mbox{d}^3\vec k$ in terms of curved Riemannian space embedded into general relativistic spacetime. This will allow us to address the problem of dynamo action in Einstein cosmology. The dynamo effect in plasmas is a competitive effect between convection of the cosmological fluid and the plasma resistivity. This is the reason why is so interesting to consider the relation between the dynamo action $\gamma$ and the plasma resistivity and its frequency. Recently, some new investigations on this directions were addressed by \cite{Rubashnyi}, by using the magnetic field correlation tensor in space of negative curvature, and by \citet{Souza} in the case of positive curvature. Also recently, \cite{Garcia} has obtained a constraint on dynamo action from COBE data, using two-dimensional spatial sections of negative curvature of Friedmann universe, based on the general relativistic magnetohydrodynamic equation derived by \citet{Marklund}.

In near future, by using the formula of \citet{Kulsrud},
$$\frac{\mbox{d}\epsilon_M}{\epsilon_M} = \gamma$$
where $\gamma$ is the magnetic field growth rate (dynamo action), we expect to compute $\gamma$ in terms of the magnetic plasma resistivity ($\eta$) and plasma frequency ($\omega_p$). In this way we shall estimate the variation of dynamo action in terms of $\eta$ and $\omega_p$.

\section{Acknowledgment}\label{thanks}

The authors would like to thank Garcia de Andrade for useful comments.

\end{document}